\documentclass[12pt,a4paper]{article}
\usepackage{graphicx}
\usepackage{times}

\title{\bf X-ray Modeling of $\eta$ Carinae \& WR140 from\\SPH Simulations}
\author{Christopher M. P. Russell$^1$, Michael F. Corcoran$^2$, Atsuo T. Okazaki$^3$,\\ Thomas I. Madura$^1$, and Stanley P. Owocki$^1$\\
\vspace{1cm}\\
\normalsize $^1$ University of Delaware, Newark, DE, USA\\
\normalsize $^2$ NASA GSFC, Greenbelt, MD, USA \\
\normalsize $^3$ Hokkai-Gakuen University, Sapporo, Japan}
\date{\mbox{}}
\begin{document}
\maketitle
\pagestyle{empty}
%
%
\def\bull{\vrule height .9ex width .8ex depth -.1ex}
\makeatletter
\def\ps@plain{\let\@mkboth\gobbletwo
\def\@oddhead{}\def\@oddfoot{\hfil\tiny\bull\quad
``The multi-wavelength view of hot, massive stars''; 39$^{\rm th}$ Li\`ege Int.\ Astroph.\ Coll., 12-16 July 2010 \quad\bull}%
\def\@evenhead{}\let\@evenfoot\@oddfoot}
\makeatother
%
%
\def\beginrefer{\section*{References}%
\begin{quotation}\mbox{}\par}
\def\refer#1\par{{\setlength{\parindent}{-\leftmargin}\indent#1\par}}
\def\endrefer{\end{quotation}}
%
%

{\noindent\small{\bf Abstract:}
The colliding wind binary (CWB) systems $\eta$ Carinae and WR140 provide unique laboratories for X-ray astrophysics. Their wind-wind collisions produce hard X-rays that have been monitored extensively by several X-ray telescopes, including RXTE. To interpret these RXTE X-ray light curves, we model the wind-wind collision using 3D smoothed particle hydrodynamics (SPH) simulations. Adiabatic simulations that account for the absorption of X-rays from an assumed point source at the apex of the wind-collision shock cone by the distorted winds can closely match the observed 2-10keV RXTE light curves of both $\eta$ Car and WR140. This point-source model can also explain the early recovery of $\eta$ Car's X-ray light curve from the 2009.0 minimum by a factor of 2-4 reduction in the mass loss rate of $\eta$ Car. Our more recent models relax the point-source approximation and account for the spatially extended emission along the wind-wind interaction shock front.  For WR140, the computed X-ray light curve again matches the RXTE observations quite well.  But for $\eta$ Car, a hot, post-periastron bubble leads to an emission level that does not match the extended X-ray minimum observed by RXTE. Initial results from incorporating radiative cooling and radiatively-driven wind acceleration via a new anti-gravity approach into the SPH code are also discussed.}
%
%
\section{Point-Source Emission Model}
\begin{figure}[h]
\centering
\includegraphics[width=17cm]{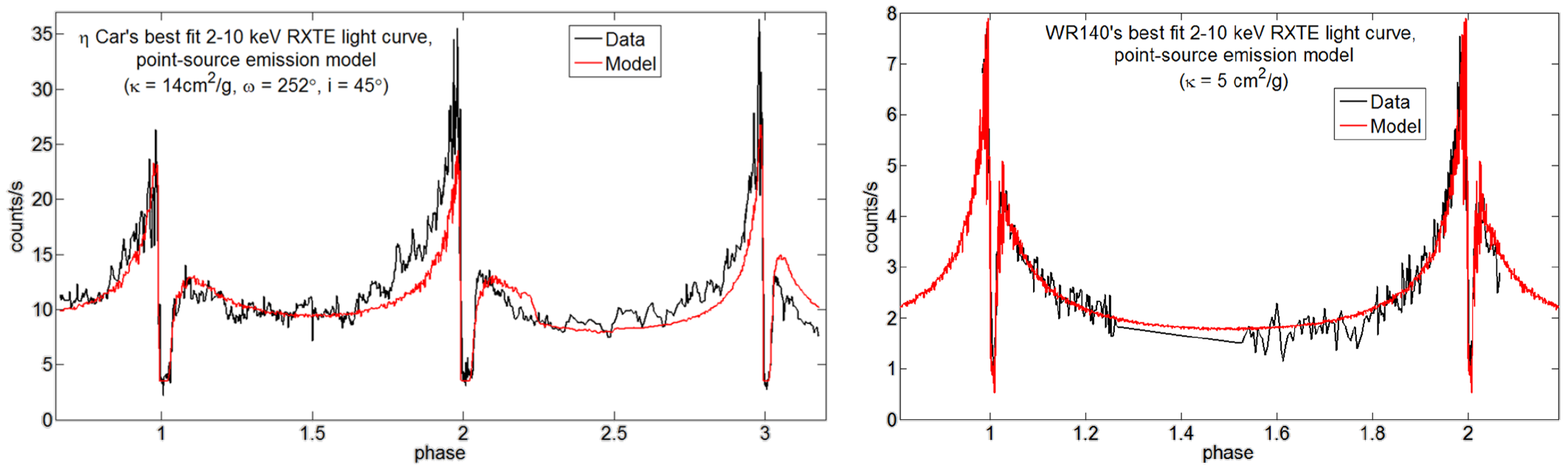}
\caption{Best-fit RXTE light curves using the point-source emission model for $\eta$ Car (left) and WR140 (right).  The model light curves are normalized to match the average maximum just before the first periastron passage (phase $\sim$0.98) and the following apastron (phase 1.5). The free parameters of the model- opacity $\kappa$, argument of periastron $\omega$, and inclination $i$- are shown for the relevant models.\label{fig_1}}
\end{figure}

$\eta$ Carinae and WR140 are long period, highly eccentric colliding wind binaries (CWBs) that provide unique laboratories for X-ray astrophysics.  Their wind-wind collisions produce hard X-rays that are strongly phase dependent; the wide range in stellar separation from the high eccentricity affects the X-ray emission, and the severe distortion in the wind geometry from significantly high orbital speeds around periastron affects the X-ray absoprtion.  Our initial attempts to model the 2-10 keV RXTE light curves of both $\eta$ Car and WR140 have applied a simple model of point-source emission plus line-of-sight wind absorption to 3D, adiabatic, smoothed particle hydrodynamics (SPH) simulations of the binary wind-wind interaction (see Okazaki et al.\ 2008 for details).  To match the recent shorter minimum of $\eta$ Car (Corcoran et al.\ 2010), the primary mass loss rate is reduced by a factor of 2.5 at phase 2.2. Many of the light curve's features are reproduced, including the shorter recent minimum.  The WR140 light curve matches remarkably well (see Fig.\,\ref{fig_1} for the light curves and Table \ref{Table1} for the parameters of the SPH simulations).

\begin{table}[h]
\caption{Parameters used in the adiabatic SPH simulations of $\eta$ Car and WR140.
\label{Table1}}
\small
\begin{center}
\begin{tabular}{| l | l | l | l |}
\hline
Parameter & $\eta$ Car & WR140 & References\\
\hline
Primary Mass (M$_{\odot}$) & 90$^{a}$ & 50$^{b}$ & a: Hillier et al. 2001\\
Secondary Mass (M$_{\odot}$) & 30$^{c}$ & 19$^{b}$ & b: Marchenko et al. 2003\\
Primary Radius (R$_{\odot}$) & 90$^{\dagger}$ & 12$^{d}$ & c: Verner et al. 2005\\
Secondary Radius (R$_{\odot}$) & 30$^{\dagger}$ & 13$^{e}$ & d: Williams et al. 1990\\
Primary Mass Loss Rate (M$_{\odot}$ yr$^{-1}$) & 2.5$\times10^{-4}$$^{f}$ & 1.2$\times$10$^{-6}$$^{g}$ & e: Pollock et al. 2005\\
Secondary Mass Loss Rate (M$_{\odot}$ yr$^{-1}$) & 1.0$\times10^{-5}$$^{f}$ & 3.8$\times$10$^{-5}$$^{g}$ &  f: Pittard and Corcoran 2002\\
Primary Wind Velocity (km s$^{-1}$) & 500$^{a}$ & 3200$^{g}$ &  g: Zhekov and Skinner 2000 (Model A)\\
Secondary Wind Velocity (km s$^{-1}$) & 3000$^{f}$ & 2860$^{g}$ & h: Corcoran 2005\\
Orbital Period (d) & 2024$^{h}$ & 2899$^{b}$ & i: Corcoran et al. 2001\\
Semi-major Axis (au) & 15.4$^{i}$ & 16.3$^{b}$ & j: Dougherty et al. 2005\\
Orbital Eccentricity & 0.9$^{i}$ & 0.88$^{b}$ & \\
Inclination (deg) & ?? & 122$^{j}$ & $^{\dagger}$: assumed M(M$_{\odot}$)/R(R$_{\odot}$) =1\\
Argument of Periastron (deg) & ?? & 46.7$^{b}$ &\\
\hline
\end{tabular}
\end{center}
\end{table}

\section{Extended Emission Model}
\begin{figure}[h]
\centering
\includegraphics[width=17cm]{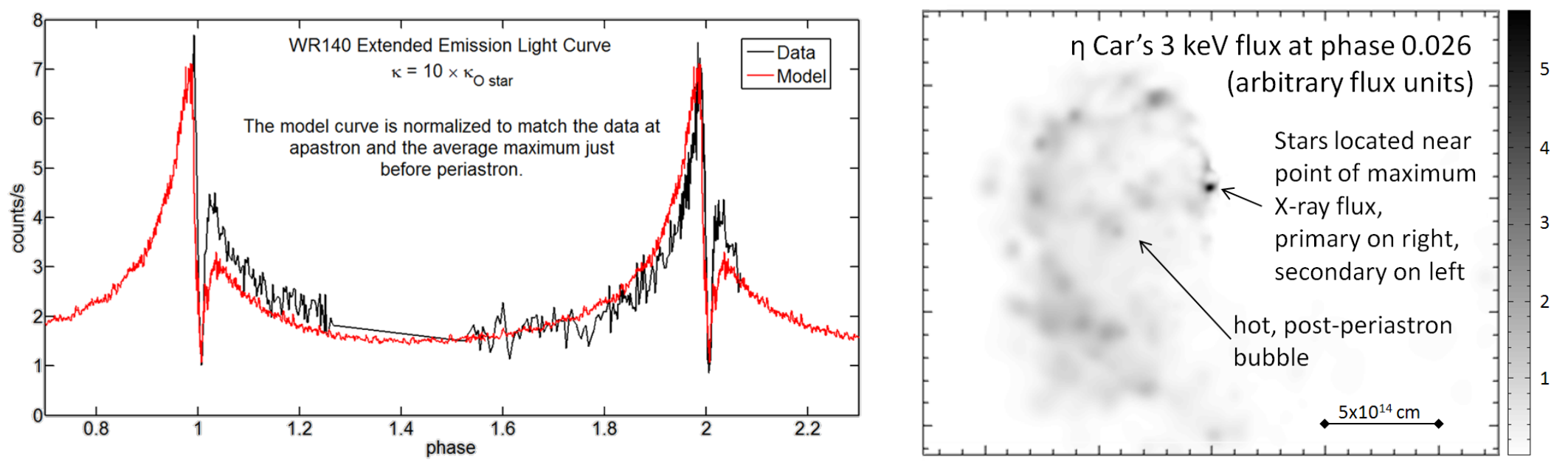}
\caption{WR140 RXTE light curve using the extended emission model (left) and the post-periastron bubble of $\eta$ Car (right).\label{fig_2}}
\end{figure}

Our more recent efforts to model the RXTE light curves of $\eta$ Car and WR140 relax the point-source approximation (e.g. Parkin et al. 2009).  The emission now originates from spatially extended post-shock gas in the wind-wind collision region according to $\rho^{2}\Lambda(E,T)$, where $\rho$ is the density and $\Lambda(E,T)$ is the emissivity as a function of energy $E$ and temperature $T$ obtained from the \textsc{mekal} code (Mewe et al.\ 1995).  The extended wind absorption is now also energy dependent.  We then use the SPH visualization program \textsc{splash} (Price 2007) to calculate the ray-tracing through the system, which generates images in various X-ray bands that combine to make a 2-10 keV X-ray light curve. Once again, the WR140 light curve matches well (assuming the opacity is 10$\times$ the opacity of an O star wind at solar abundances, an assumption that will be relaxed in future work).  The same is not true for $\eta$ Car, however, where a hot, post-periastron bubble blown into the slow, dense primary wind by the much faster companion wind prevents the reproduction of the extended X-ray minimum (see Fig.\,\ref{fig_2}).

\begin{figure}[h]
\centering
\includegraphics[width=17cm]{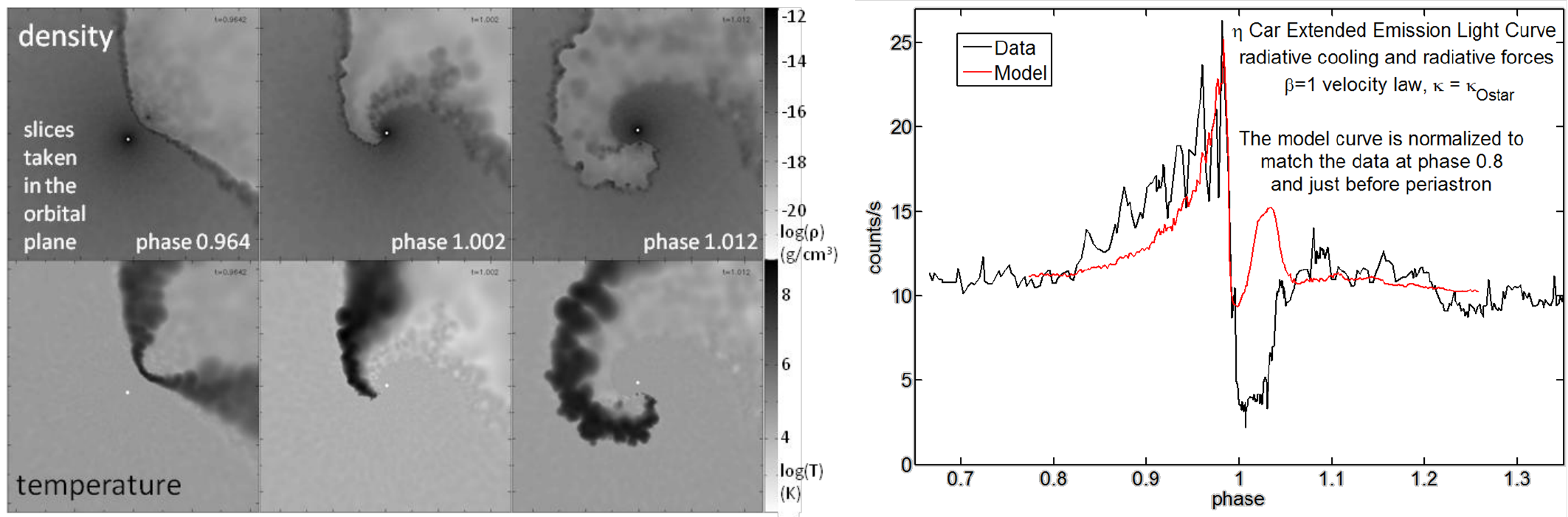}
\caption{Density and temperature snapshots of an SPH simulation of $\eta$ Car that includes radiative forces and radiative cooling (left) and the resulting RXTE model light curve using the extended emission model (right). In the SPH snapshots, $\eta$ Car is the white dot, while the companion is shown by the black dot in the left density panel and then orbits counterclockwise. The snapshots show that hot material exists between the stars before periastron (left panels), vanishes at periastron (center panels) from radiative forces and radiative cooling, and begins to reappear after periastron (right panels).  The right panels also shows the hot, post-periastron bubble that forms below both stars, which prevents the matching of the RXTE light curve during the X-ray minimum.  \label{fig_3}}
\end{figure}

Radiative cooling, via the Exact Integration Scheme (Townsend 2009), and the radiative driving of the stellar winds, via an anti-gravity approximation (an assumed outward acceleration that competes with gravity instead of the full Castor-Abbott-Klein calculation), have been implemented.  The acceleration of the secondary wind towards the primary star drastically decreases around periastron due to radiative inhibition, where the radiation from one star provides enough force on it's companion's wind to inhibit it's acceleration (Stevens \& Pollock 1994).  This leads to the secondary's wind colliding with the primary's wind at a much lower speed directly between the stars, so the high temperature shock cone of $\eta$ Car collapses, as was proposed by Parkin et al.\ (2009).  However, there is still a high temperature shock on the backside of the secondary, the side opposite the primary star.  The high orbital speed at periastron essentially causes the secondary to become buried in the primary's wind, so the secondary wind on the backside will collide with primary wind material.  Since the acceleration of the backside's secondary wind is not affected by radiative inhibition, this portion of the wind still collides at a high speed (although the collision is co-moving instead of head-on) to create a high temperature, post-periastron bubble.  The hard X-rays produced in this post-periastron bubble prevent the reproduction of the minimum of the RXTE light curve with the extended emission model.  Outside the minimum, the model matches fairly well (see Fig.\,\ref{fig_3}).

\section{Conclusions \& Future Work}
The results of using 3D SPH simulations to model the wind-wind interaction of the CWBs $\eta$ Carinae and WR140 have been presented, as well as our attempts to model these systems' RXTE light curves.  Although the point-source emission model produces remarkably good results for both systems when applied to the adiabatic SPH simulations, this model is clearly an approximation that needs to be relaxed.  Upon doing this by implementing an extended emission model, the resultant X-ray light curve for WR140 matches pretty well, while $\eta$ Car's light curve does not match well during the X-ray minimum.  A hot, post-periastron bubble forms in $\eta$ Car from the fast companion wind blowing into the slow primary wind just after periastron, and this large emitting region creates a significant digress from the point-source result.  On the other hand, the similar wind speeds in WR140 prevent a post-periastron bubble from forming, so the emission is approximately point-like throughout the orbit, and hence in good agreement with the point-source emission model results.

To improve the match of the extended emission model during $\eta$ Car's RXTE X-ray minimum, radiative cooling and radiative driving of the winds via an anti-gravity approach were added to the SPH code (previous SPH simulations were adiabatic and had the winds leave their star at terminal speed).  While this leads to the collapse of the high temperature shock between the stars at periastron due to radiative inhibition, it does not affect the hot gas in the remainder of the post-periastron bubble.  Therefore, there is still excess X-ray emission at periastron compared to the RXTE observations.

At the time of this conference, the results from the SPH simulations with radiative cooling and radiative driving are relatively new, so our immediate future work will be to further explore and understand these results.  Of particular interest is the post-periastron bubble, which appears to be preventing the matching of the RXTE light curve during the X-ray minimum.  Furthering the intrigue about this bubble are the results of Parkin et al.\ (these proceedings) where a 3D grid-based hydrodynamics code was used to model $\eta$ Car, and although the density and temperature structures compare very well from the grid-based code to the SPH code, and they too have excess emission during the X-ray minimum, the authors do not claim to see a post-periastron bubble.  Other future work will include modeling WR140 with the radiative cooling and radiative driving improvements to the SPH code, modeling other CWBs such as HD5980, and improving the resolution of the extended emission model to begin modeling the X-ray spectra of CWBs.

%
%
\section*{Acknowledgements}
CMPR and TIM acknowledge support from NASA GSRP Fellowships.
%
%
\footnotesize
\beginrefer

\refer Corcoran M.F., 2005, AJ, 129, 2018

\refer Corcoran M.F., Hamaguchi K., Pittard J.M., Russell C.M.P., Owocki S.P., Parkin E.R., Okazaki A.T., 2010, ApJ, 725, 1528

\refer Corcoran M.F., Ishibashi K., Swank J.H., Petre R., 2001, ApJ, 547, 1034

\refer Dougherty S.M., Beasley A.J., Claussen M.J., Zauderer B.A., Bolingbroke N.J., 2005, ApJ, 623, 447


\refer Hillier D.J., Davidson K., Ishibashi K., Gull T., 2001, ApJ, 553, 837

\refer Marchenko S.V., Moffat A.J., Ballereau D., et al., 2003, ApJ, 596, 1295

\refer Mewe R., Kaastra J.S., Liedahl D.A., 1995, Legacy, 6, 16

\refer Okazaki A.T., Owocki S.P., Russell C.M.P., Corcoran M.F., 2008, MNRAS, 288, L39

\refer Parkin E.R., Pittard J.M., Corcoran M.F., Hamaguchi K., Stevens I.R., 2009, MNRAS, 394, 1758

\refer Parkin E.R., et al. 2010, these proceedings

\refer Pittard J.M., Corcoran M.F., 2002, A\&A, 383, 636

\refer Pollock A.M.T., Corcoran M.F., Stevens I.R., Williams P.M., 2005, ApJ, 629, 482

\refer Price D.J., 2007, PASA, 24, 159

\refer Stevens I.R., Pollock A.M.T., 1994, MNRAS, 269, 226

\refer Townsend R.H.D., 2009, ApJS, 181, 391

\refer Verner E., Bruhweiler F., Gull T., 2005, ApJ, 624, 973

\refer Williams P.M., van der Hucht K.A., Pollock A.M.T., Florkowski D.R., van der Woerd H., Wamsteker W.M., 1990, MNRAS, 243, 662

\refer Zhekov S.A., Skinner S.L., 2000, ApJ, 538, 808\\

\endrefer

\end{document}